# The role of intermolecular interactions in stabilizing the structure of the nematic twist-bend phase

Katarzyna Merkel[1], Barbara Loska[1], Chris Welch[2], Georg H. Mehl[2], Antoni Kocot[1]

[1]Institute of Materials Engineering, University of Silesia, 75 Pułku Piechoty 1A, 41-500 Chorzów, Poland
[2]Department of Chemistry, University of Hull, Hull HU6 7RX, UK

**Abstract**

The relationship between the molecular structure and the formation of the $N_{TB}$ phase is still at an early stage of development. This is mainly related to molecular geometry, while the correlation between the $N_{TB}$ phase and the electronic structure is ambiguous. To explore the electronic effect on properties and stabilization of the $N_{TB}$ phase we investigated 2′,3′-difluoro-4,4′′-dipentyl-p-terphenyl dimers (DTC5Cn). We used IR polarized spectroscopy, which can at least in principle, bring information about the ordering in $N_{TB}$ phase. All dimers show a significant drop of the average value of the transition dipole moment $d\mu/dQ$ for parallel dipoles at the transition to the $N_{TB}$ phase, and an increase for perpendicular dipoles, despite its remaining unchanged for the monomer. These results coincide well with DFT simulations of vibrational dipole derivatives for molecules assembled in pseudo-layers of the $N_{TB}$ phase. The DFT calculations were used to determine the geometric and electronic properties of the hydrogen bonded complexes. We have provided experimental and theoretical evidence of stabilization of the $N_{TB}$ phase by arrays of multiple hydrogen bonds (XF····HX, X-benzene ring).

**Introduction**

The structure-property relationship for the LC dimers has been extensively studied; leading to the common belief that molecular curvature is necessary for the formation of the $N_{TB}$ phase, and its stabilization increases as the molecular bending angle decreases, defined as the angle between the two mesogenic arms [1-4]. In addition to molecular bending, several reports have suggested that intramolecular torsion, conformational distribution, bending angle fluctuation, and the effect of free volume can affect the transition temperature $T_{NTB-N}$ and can also influence stability of the $N_{TB}$ phase. The key observation of recent experiments is that on cooling the dimers through the $N$-$N_{TB}$ phase transition the correlation length of the spatial periodicity drops, while the heliconical orientational order becomes more correlated [5]. Despite these efforts, the relationship between the molecular structure and the formation of the $N_{TB}$ phase is still at an early stage of development. This is mainly related to molecular geometry, while the correlation between the $N_{TB}$ phase and the electronic structure of the molecules responsible for phase formation is ambiguous. The geometry of helices and their close-packed structures it is very essential to characterize the geometric properties of helices traced by atoms in liquid crystalline molecules because their function is driven by structure and dynamics.

Recently, there have been several reports on the induction of the twist-bend phase ($N_{TB}$) in LCs dimers using hydrogen bonds [6-8]. Intermolecular hydrogen bridges have a direct impact on the structure and durability of the molecular network. This is related to the energy of the system as well as to the configuration of the periodic arrangement of the molecules. The role of hydrogen bonding in the formation of liquid crystallinity in mixtures containing bipyridines and 4-pentoxybenzoic acid (supramolecular aggregates) has been well investigated using Furrier Infrared Spectroscopy (FTIR) [9-11].

Hydrogen bonding interactions play an important role in many chemical and biological systems. Fluorine acting as a hydrogen-bond acceptor in intermolecular and intramolecular interactions has been the subject of many controversial discussions and there are differing opinions about it [12-14]. We have proposed a correlation between the propensity of fluorine to be involved in hydrogen bonds and its 19F NMR chemical shift. We now provide additional experimental and computational evidence for this correlation [15-18]. The incorporation of one or more fluorine moieties in a molecule can dramatically change its properties through a direct and/or indirect effect of the fluorine atoms [19, 20].

**Experimental**

**Materials**

To explore the electronic effect on properties and stabilization of the $N_{TB}$ phase and a delicate interplay between the $N_{TB}$ structure and the strength of the intermolecular interaction we investigated bent-shaped dimers containing double fluorinated terphenyl moiety (Fig. 1). In the present paper we used IR polarized spectroscopy to measure all absorbance component of the homologous series of difluoroterphenyl dimers DTC5Cn, with odd numbers of carbons in the linkage, n=5,7,9,11. Detailed synthesis is presented in the paper *W.D. Stevenson et.al* [2, ESI]. The sample for the IR study was aligned in between the two optically polished ZnSe windows. In order to obtain the homogeneous orientation of molecules, windows were spin coated with a SE-130 commercial polymer aligning (Nissan Chemical Industries, Ltd). The cells were assembled with parallel arrangement of the rubbing direction. In order to obtain the homeotropic alignment of samples we used a commercial solution of the AL 60702 polymer (JSR Korea). Mylar foil was used as a spacer and thickness of cells fabricated was determined to be in the range from 5.1 – 5.6 µm, by the measurements on the interference fringes using a spectrometer interfaced with a PC (Avaspec-2048). The sample was capillary filled by heating the empty cell in the nematic phase, five degrees below the transition to the isotropic phase. The quality of the alignment has been tested using polarizing microscopy. The texture of the sample was monitored using a polarizing microscope that was used for identifying the phase prior to its investigation by polarized IR spectroscopy.

**Infrared spectroscopy**

The infrared spectra were recorded using an Agilent Cary 670 FTIR spectrometer with a resolution of 1 cm$^{-1}$ and these spectra are averaged over 32 scans. The experiment was performed using transmission method with polarized IR beam (Fig. 1b). An IR-KRS5 grid polarizer is used to polarize the IR beam. IR spectra have been measured as a function of the polarizer rotation angle in the range 500-4000 cm$^{-1}$ of wavenumbers. Measurements were performed on slow cooling and heating at the rate of 0.1 K/min. Temperature of the sample was stabilized using PID temperature controller within ±2 mK.

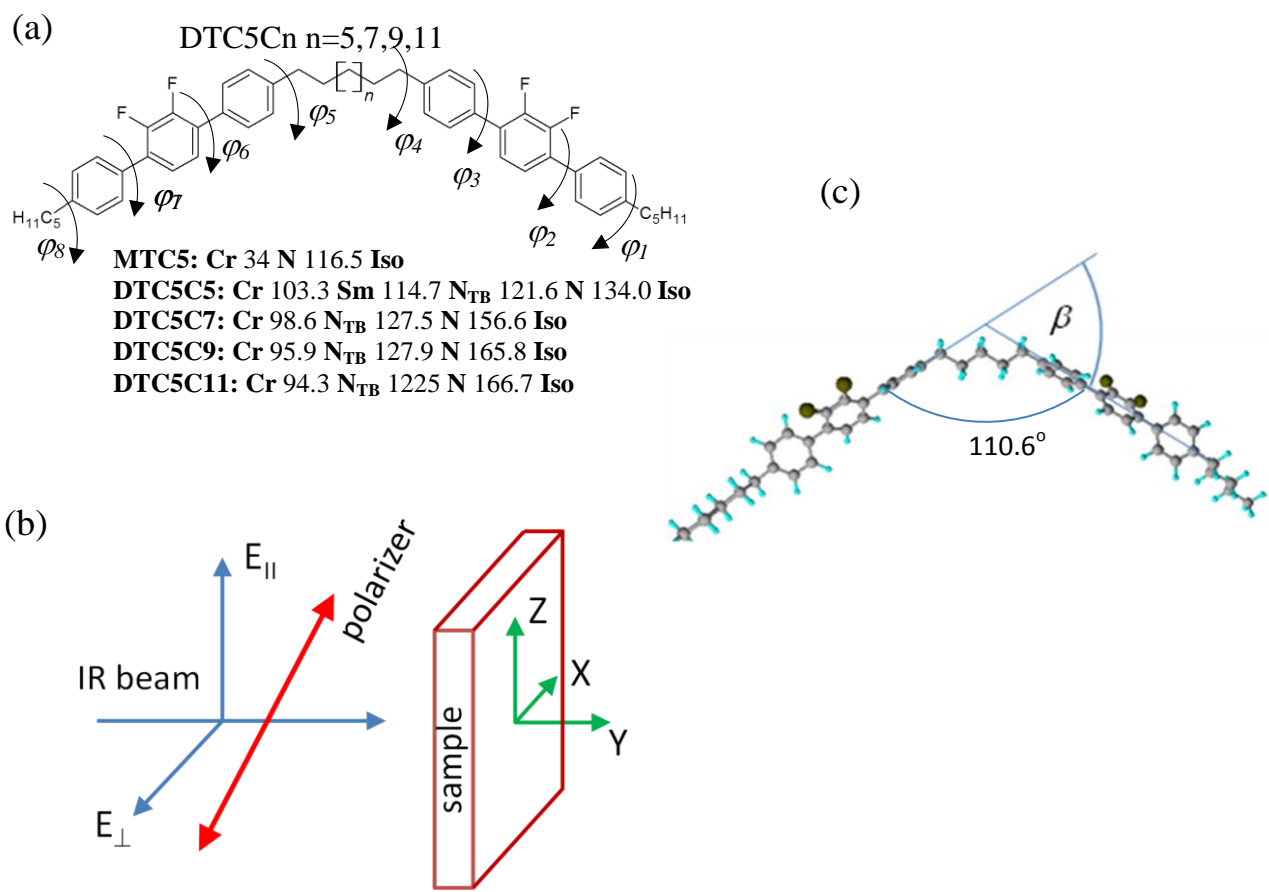

Fig.1. 2′,3′-difluoro-4,4′′-dipentyl-p-terphenyl dimers (DTC5Cn) (n=5,7,9,11). (a) Transition temperature and molecular structure of DTC5Cn. (b) Schematic of polarized infrared transmission technique at normal incidence of light. (c) Simulated structure of the helical conformer of DTC5C5 molecule. $\beta$ – angle that the core makes with the bow string axis of the dimer.

In an orientationally ordered material, the absorbance components are dependent on the angle between the alignment axis and the polarization direction of the incident beam. At a microscopic level, the infrared absorption depends on the angle between the molecular transition dipole moment $\mu_i$ of the particular absorption band and the polarization of the IR beam. The average IR absorbance $A_0=(A_X+A_Y+A_Z)/3$, of the particular vibrational modes are determined by how the electric dipole moment of the system changes with the atomic oscillations. To the lowest order, the required quantities are proportional to the derivatives of the dipole moment with respect to the vibrational normal modes, i, of the system, evaluated at the equilibrium geometry. The IR absorbance of the $i_{th}$ vibrational mode is given by [21]

$$A_i = \int_{\nu 1}^{\nu 2} A(\nu) d\nu = \frac{N\pi}{3c}\left[\frac{d\mu_i}{dQ_i}\right]^2 \qquad (1)$$

where: N is the number of molecules per unit volume, μ is the molecule dipole moment, and Qi is the normal coordinate corresponding to the $i_{th}$ mode.

For an anisotropic system the use of vibration spectroscopy can provide information about the orientational order of individual functional groups of molecules, and also about specific intramolecular and intermolecular interactions of these groups. In the LC phases the absorbance components of the selected vibrational band are related to the orientation of corresponding transition dipole moments.

$$A_{X,Y} = A_0 - B\left[\tfrac{1}{3}S\left\{(\mu_i)_n^2 - \tfrac{1}{2}((\mu_i)_l^2 + (\mu_i)_m^2)\right\} + \tfrac{1}{6}D\left\{(\mu_i)_l^2 - (\mu_i)_m^2\right\}\right]$$

$$A_Z = A_0 + B\left[\tfrac{2}{3}S\left\{(\mu_i)_n^2 - \tfrac{1}{2}((\mu_i)_l^2 + (\mu_i)_m^2)\right\} + \tfrac{1}{3}D\left\{((\mu_i)_l^2 - (\mu_i)_m^2)\right\}\right]$$

(2)

Where: $S$ and $D$, are orientational order parameters, of the long axis and molecular biaxiality, respectively, in the uniaxial nematic phase. $S = S_{zz}^Z$, $D = S_{xx}^Z - S_{yy}^Z$.

Following the Saupe ordering matrix [22,23], the parameter S is a measure of the increase in compatibility of the molecule long axis with a nematic director while D describes the rotational biasing of the short molecular axis. Assuming the transition dipoles remain constant in the temperature range of liquid and LC phase, if the correlations of the transition dipoles are neglected, then by the summation rule, $B\mu^2 = A_0$.

**DFT calculations**

All calculations in this study were performed using the Gaussian09 program, version E.01 [24]. The molecular structures, the binding energy of no-specific hydrogen bond, the harmonic vibrational force constants, and absolute IR intensities were calculated using a density functional theory (DFT) with the Becke's three-parameter exchange functional in combination with the Lee, Yang, and Parr correlation functional the B3-LYP method with the diffusion basis set: 6-311+G [25,26]. In order to find the most stable conformation of the dimer the optimization of the geometry were performed in a few steps. All possible conformation of dimers we have considered is defined by the values of dihedral angles $\varphi_1$ - $\varphi_8$ (Fig 1a). In the first stage the energy barriers for the internal rotation of the terphenyl (torsional angles $\varphi_2$ and $\varphi_3$) have been determined. In a further step, energy barriers for the rotation around the dihedral angle ($\varphi_1$, $\varphi_4$) between terphenyl and the linker / tail in the MTC5 monomer were determined. The approximate potential energy functions have been calculated at intervals of 10. In the calculations, the torsional angles ($\varphi_1$-$\varphi_4$, each in turn) fixed at arbitrary selected values while the other geometrical parameters were optimized; relaxed potential energy scans were performed. This procedure allowed determining the values of torsion angles for which the minimum energy was obtained. As the energy barrier for the internal rotation in the alkyl chain is very small (approx. 1 kJ/mol), therefore, we did not consider any other spacer and tail conformations than the all-trans. In ordered phases, the alkyl chains can in principle have all possible conformations, therefore adopting all he all-trans conformation to be a better representation of the average molecular shape of dimers [27]. In the next step, taking into account the determined values of torsion angles, full optimization of the geometry was performed for DTCnC dimers with a value of n = 5.7. All DFT optimization were carried out with the following convergence criteria used with Berny algorithm (all values in atomic units): the maximum component of the force was set to 0.00045, the root-mean square (RMS) of the forces calculated for the next step—smaller then 0.0003—the computed displacement for the next step—smaller than 0.0018—and the RMS of the displacement below 0.0012. These criteria restrict the dependence of the final geometry parameters on the initial starting geometry. Therefore, the vibrational calculations were done for the most probable conformers with the lowest energy for the MTC, the DTC5C5 and the DTC5C7. For B3LYP frequency calculations, a pruned 99,590 grid was used to obtain a more accurate numerical integration; it is important for computing low frequency modes. Maximal force (in atomic units) that was lower than 3.6x $10^{-5}$ after geometry optimization. The rotational frequencies were very close to zero, the translational frequencies smaller than 7.5 cm$^{-1}$. The theoretical vibrational frequencies were scaled by one coefficient equal 0.98 in order to simplify the comparison with experiment.

### Results of the experiment

#### Molecular structure and vibrations

The possible structures of the each arm of the dimer can be classified by the torsion angles: $\varphi_1$ - $\varphi_8$ (see Fig. 1a). The rotation around the inter-ring of the terphenyl it showed the two most energy-favorable conformers: helical ($\varphi_2$=-40, $\varphi_3$=-40) and twisted ($\varphi_2$=40, $\varphi_3$=40). On the other hand, the rotation around the bond C-C-C-C between terphenyl and the linker or tail defined by $\varphi_1$ and $\varphi_4$ angles showed a minimum energy for value of 90°. For the fully optimized geometry for the DTC5C5 dimer, we obtained the torsion angle values: $\varphi_1$ =-90.25, $\varphi_2$=-43.7, $\varphi_3$=-43.5, $\varphi_4$=93.4, $\varphi_5$ = 85.3, $\varphi_6$=-43.5, $\varphi_7$=-43.7, $\varphi_8$=-90.25 for helical conformer and $\varphi_1$ = -89.2, $\varphi_2$=-42.7, $\varphi_3$=43.5, $\varphi_4$=-84.5, $\varphi_5$ = 95.6, $\varphi_6$=-42.8, $\varphi_7$=43.5, $\varphi_8$=-90.93 for twisted conformer. The value of the $\varphi_4$/$\varphi_5$ angle is the most crucial for determining the bend angle of the molecule. In this case the bending angle of the molecule was determined 110.6 degrees.

All of the observed bands in the $N_{TB}$ phase have been explained by the coexistence of the helical and the twisted conformers. Most of the observed bands are assigned to the overlaps of the bands attributable to the helical and the twisted conformers. Most changes in the FTIR spectra associated with different terphenyl conformation concern bands in the wavenumber range from 500-800 cm$^{-1}$. The infrared bands observed at 1317, 825, 750, 660, 560 are assigned to the helical conformer. The infrared bands at 485 cm$^{-1}$, 630 cm$^{-1}$, 645 cm$^{-1}$ and 800 cm$^{-1}$ are attributed to the twisted conformer. The calculated FTIR spectra of the twisted and the helical conformers of monomer (MTC5) are shown in comparison with the experimental spectra in the nematic phase of the MTC5 in Fig. 2a

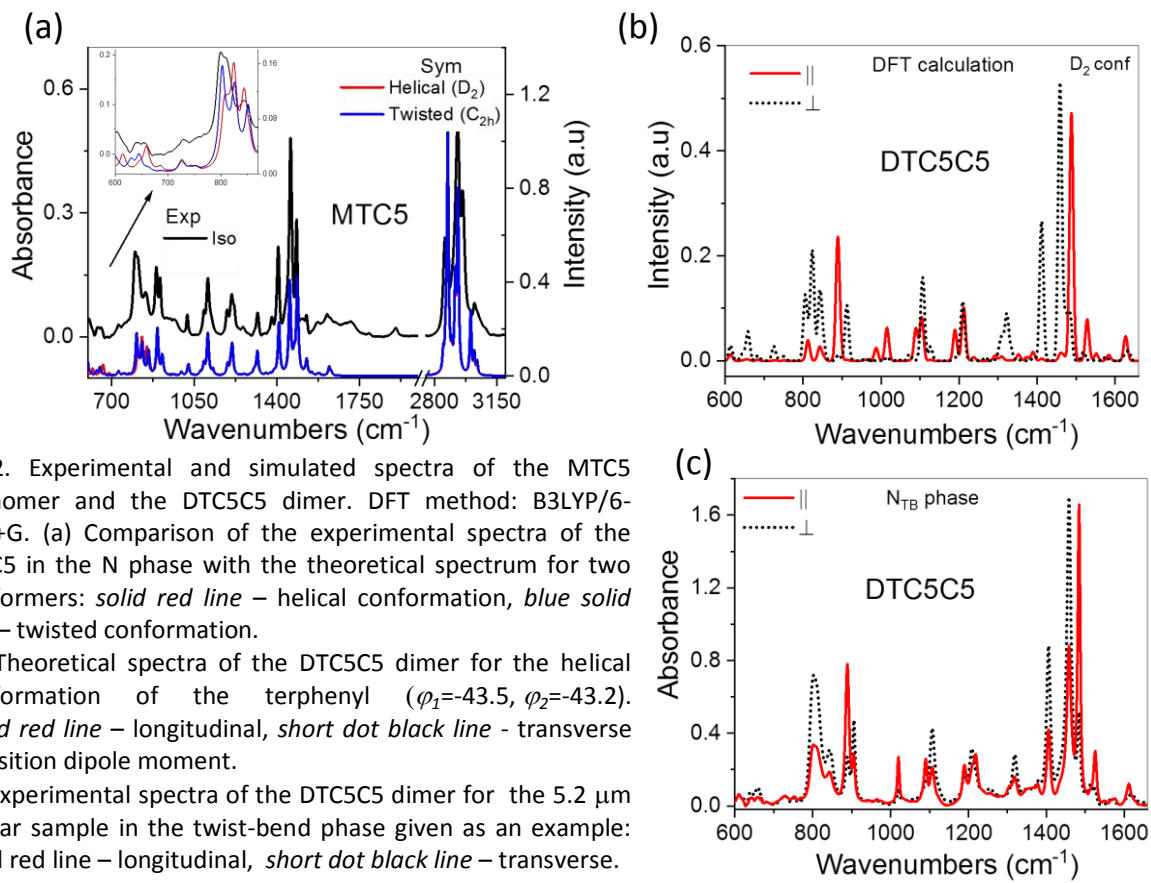

Fig.2. Experimental and simulated spectra of the MTC5 monomer and the DTC5C5 dimer. DFT method: B3LYP/6-311+G. (a) Comparison of the experimental spectra of the MTC5 in the N phase with the theoretical spectrum for two conformers: *solid red line* – helical conformation, *blue solid line* – twisted conformation.
(b) Theoretical spectra of the DTC5C5 dimer for the helical conformation of the terphenyl ($\varphi_1$=-43.5, $\varphi_2$=-43.2). *Solid red line* – longitudinal, *short dot black line* - transverse transition dipole moment.
(c) Experimental spectra of the DTC5C5 dimer for the 5.2 μm planar sample in the twist-bend phase given as an example: solid red line – longitudinal, *short dot black line* – transverse.

For better identification of main frequencies in the dimers, we compared the experimental data for dimers with the theoretical infrared spectra for longitudinal and transverse transition dipole moment. Most of the fundamentals in the range of 500–1700 cm$^{-1}$ were very well reproduced by vibrations in the experimental spectra, except the bands in the range of 2900 – 3100 cm$^{-1}$, mainly because of the strong anharmonicity. Figure 2b and 2c show comparison of the simulated spectra and experimental one for the DTC5C5 dimer as an example.

**Infrared absorbencies**

Using polarized FTIR spectroscopy we can directly analyze the temperature dependence of the absorbance components in the temperature range of the nematic and the $N_{TB}$ phases. As it can be predicted from the S parameter we can expect distinctly different behavior of the absorbance components for the bands that have transition dipole longitudinal and transverse with respect to the core axis. By combining the FT IR results for homogenous planar and homeotropic alignment of the sample we can obtain all ($A_X$, $A_Y$, $A_Z$) components of the IR intensities. The average intensity $A_0$ and the related transition dipoles were analyzed for the series DTC5Cn in the temperature range of the N and the $N_{TB}$ phases and for MTC5 monomer in the *N* phase as a reference. Several vibrational bands are selected to be analyzed in the mid FTIR range that offers significant dichroism of the band. These are the phenyl stretching band ($vCC$) at wavenumbers: 1485, 1460 and 1406 cm$^{-1}$. The combinational band at 905 cm$^{-1}$ that can be assigned to the phenyl in plane stretching with symmetric C-F stretching ($vCC+v_sCF$) and phenyl in plane deformation with asymmetric C-F stretching at 890 cm$^{-1}$ ($\beta CC + v_{as}CF$). Band assignments in the FTIR spectra were made based on simulated spectra of the monomer (MTC5) and the DTC5C5 dimer (Fig 1).

Figure 3 shows temperature dependence of absorbances for the bands at 890 cm$^{-1}$ ($\beta CC + v_{as}CF$) and 905 cm$^{-1}$ ($vCC+v_sCF$), which involve mostly the C-F bond of the aromatic cores. They correspond to the longitudinal transition dipole and transverse transition dipole, respectively. For the longitudinal dipole ($\alpha$=0) the $A_z$ component is initially growing in the N phase, then starts to decrease in the $N_{TB}$ phase for the dimer DTC5C7 (Fig.3). The trend is expected to be opposite for both perpendicular components, $A_X$ and $A_Y$, but there is not much change observed on entering the $N_{TB}$ phase. In this context it is interesting to compare the absorbance dependence for the monomer MTC5, which has the same aromatic core structures as the dimer DTC5C7. In order to do so, we need to consider the angle, $\beta_2$, that the core makes with the bow string axis of the dimer. The angle is expected to be $\beta/2$=34.7° for an all *trans* conformation of the dimer linkage as found from DFT calculations but in the higher temperature range the other conformers can also contribute significantly (see Fig.1c). The corresponding order parameter of the monomer scaled for dimer can be found as a product:

$$S_{DIM} = S_{MON} P_2(\cos(\beta_2)) \qquad (3)$$

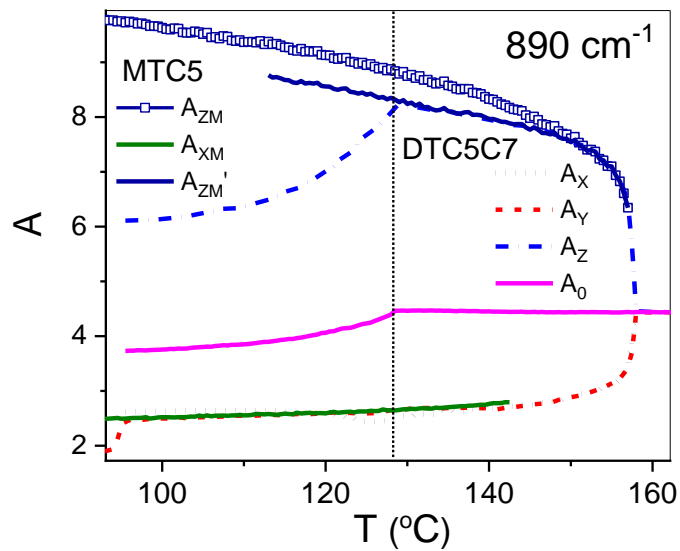

Fig.3. Normalized IR absorbance vs temperature behavior for the planar and homeotropic orientations of the monomer MTC5 and the DTC5C7dimer for the 890 cm$^{-1}$ vibration ($\beta CC + \nu_{as}CF$). Longitudinal $A_Z$ and transverse $A_X$ components for the planar sample; $A_Y$ absorbance for the homeotropic. $A_0$ - average absorbance. Z is fixed parallel to the optical axis in the N phase. $A_{ZM}$ - component of the monomer scaled to $A_Z$ component of the DTC5C7 dimer by an angle $\beta_2 \cong 30°$, $A_{ZM}'$ – as above but the angle $\beta_2$ is gradually increasing from 30° to 34.7°.

The best accordance of the $A_{ZM}$ component of the monomer scaled to the $A_Z$ component of the DTC5C7 dimer, in the range of nematic phase, can be obtained for the angle $\beta_2$ that is gradually increasing from 30° to 34.7°, the latter angle as expected from DFT simulation (Fig. 1c). The other reason for the later inconsistence might originate from the increasing of the molecular tilt with respect to the ordering axis that preceding the $N_{TB}$ phase. The case of the transverse dipole, Fig. 4, is more complex since the parameter $D$ can significantly contribute to the formula for the absorbance components. Nevertheless, for small values of the $D$ – order parameter [28] the behavior is just opposite to the case of longitudinal dipole (Fig.3) for the dimer DTC5C7 (Fig. 4). Here again, a significant change of the trend for the Az component of the absorbance can be observed while the temperature behavior of the components $A_X$, $A_Y$ do not show a significant change when observed on entering the $N_{TB}$ phase. On that basis we found a more general reason for the peculiar behavior of the spatial components. This is coming from summation rules. It is usually assumed, that the transition dipoles are interacting independently with the field of IR beam and therefore they are commonly not temperature dependent. As a consequence of the eq. (1), the average absorbance depends only on the mass density of the sample. Indeed the sum rule of the absorbance components for the monomer MDT5 proves that transition dipoles $\mu_l$ for longitudinal dipole remain constant within the whole temperature range of the nematic phase. For transverse dipoles, however, the dependence is steeper than expected from density dependence, Fig.4. This is likely due to increasing dipole-dipole correlation on cooling. If this is compared with the results for the DTC5C7 dimer, as shown in Figures 3 and 4 we can see that transition dipoles show a good coincidence with the monomer dipole within the nematic phase.

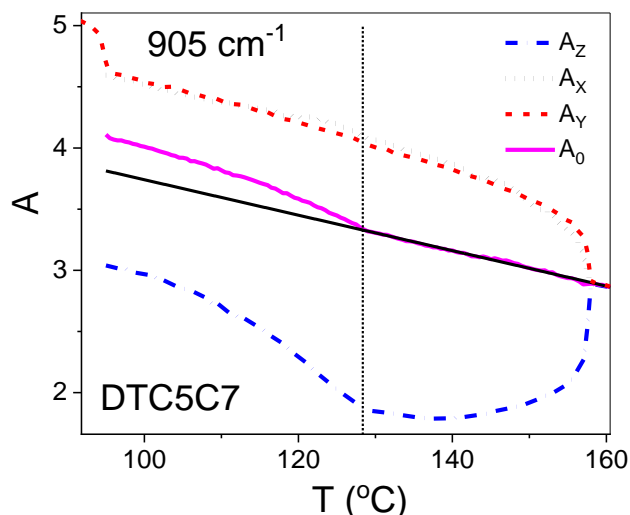

Fig.4. Normalized IR absorbance vs temperature behavior for the planar and homeotropic orientations of the DTC5C7 dimer for the 905 cm$^{-1}$ vibration ($\beta CC + \nu_s CF$). Longitudinal $A_Z$ and transverse $A_X$ components for the planar sample; $A_X$ absorbance for the homeotropic. $A_0$- average absorbance. Z is fixed parallel to the ordering direction.

However in the temperature range of the $N_{TB}$ phase there is a strong inconsistency between the temperature behavior of parallel $A_z$ and perpendicular $A_x$, $A_y$ absorbance components for both types of bands: the longitudinal and transverse dipole. In the $N_{TB}$ phase, they decline significantly from their trends in the nematic phase. For the longitudinal dipoles at the transition to the $N_{TB}$ phase all dimers show a significant drop of the average absorbance, $A_0$, and thus also $d\mu/dQ$. Both absorbance and dipole are increased for transverse dipoles. This is in contrast to the observations for the monomer where those values remain unchanged. This observation is consistent for all homologues in the series of DTC5Cn dimers.

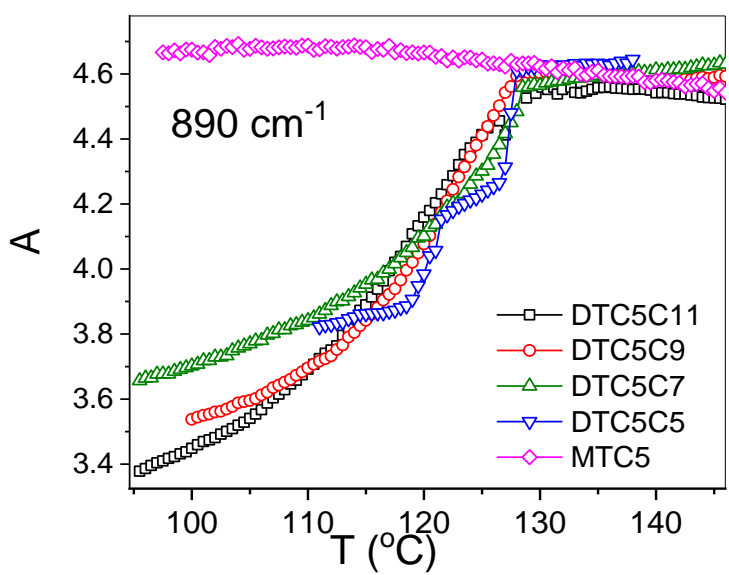

Fig.5. The average absorbance vs temperature behavior of dimers for the 890 cm$^{-1}$ vibration.

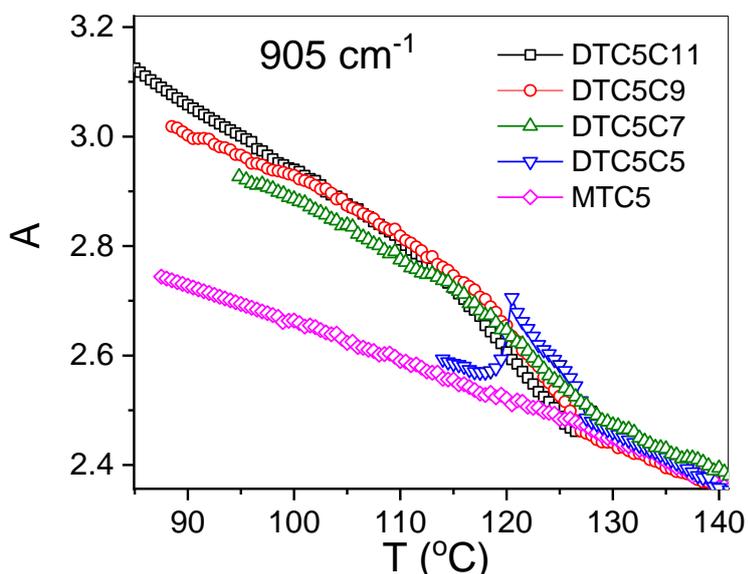

Fig.6. The average absorbance vs temperature behavior of the dimers for the 905 cm$^{-1}$ vibration.

Analyzing the results of temperature dependencies of the average absorbance $A_0$, we can conclude that for the monomer it behaves quite typically, i.e. depends mainly on the material density. Meanwhile, for the dimers it is clear that at the transition to the $N_{TB}$ phase, the average absorbance certainly cannot be attributed to the change of the density. The primary reason for this behavior could be the change in the arrangement going from the nematic to the twist-bend phase. This behavior is due to the tight and specific packing of molecules in the $N_{TB}$ phase. There is also a tendency in the $N_{TB}$ phase to form separate domains of the same sense of chirality. The helixes of the same sense like to pack themselves in a "*zipper-like*" way, within one domain or supramolecular fibers [29]. As we recently reported for the DTCnC dimers [28], we observed a sudden increase of the molecular biaxiality order parameter in the $N_{TB}$ phase, while in the $N$ phase it was negligible. These finding motivate us to analyze the possible intermolecular interactions that lead to the bond orientation. In such kind of arrangement it might be possible to select the specific type of interactions that may change the transition dipole moment of the vibrational bands. For the further analysis we shall neglect all such interactions which are already present in the nematic phase of the monomer. We also used DFT modelling to analyze possible intermolecular interactions that would lead to bond orientation and thus stabilization of the twist-band phase.

**Hypothetical arrangement of DTCnC molecules formed by intermolecular weak Interactions – DFT modeling**

In order to confirm our assumptions about the geometry of the helix and its close-packed structures, in which weak intermolecular interactions of the hydrogen bond type may play a significant role, we optimized the geometry for several systems of two interacting molecules (see Fig. 7). To reduce the computation time, we only considered one dimer arm in the interaction pairs (MTC5). In the first step the global minimum structure of single molecule was carried out as described in the section "molecular structure and vibrations".

The input file for the calculations contained optimized structures, both in twist or helical conformation, where the distance of the closest atom of one molecule was greater than 4 Å from the nearest atom in second molecule. In these input files, both interacting molecules were positioned exactly parallel to each other with *C-F* bond oriented in the same direction like in Fig 7. (1) or opposite (towards each other) like in Fig.7 (2). Then one of the molecule was shifted parallel to the other to avoid them stacking in the local minima (3). For three different arrangements of the interacting molecules, full optimization with the vibrational frequency calculation was performed using B3LYP/6-311+G. The determination of the interaction energy for a weakly bonded system is mostly estimated with the use of the so-called supermolecular approach, whereby the interaction energy difference between the whole system and its subunits is given. Therefore, in order to accurately determine the interaction energy, calculations have been performed by the supermolecular method using the basis set superposition error (BSSE) correction by the counterpoise (CP) method [30-32].

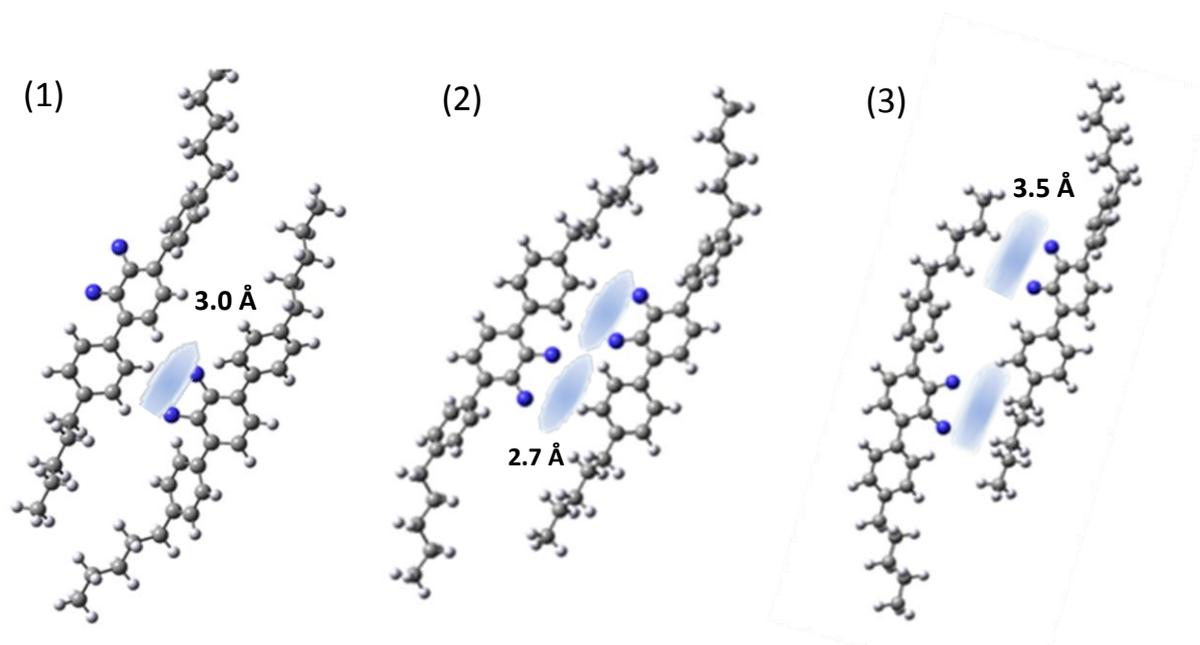

Fig. 7. Domain structures of MTC5 molecules (helical conformer) formed by intermolecular hydrogen bonds after optimization using B3LYP/6-311+G method.

The cohesive energy of the X-H⋯F-X hydrogen bonding has been estimated to be between 4.5 and 8.5 kJ/mol depending on arrangements. The largest energy of cohesion, (8.5 kJ/mol), is obtained for the system (2) with two pairs of H⋯F hydrogen bond. For the case (1), with the one pair of H⋯F bond, the binding energy become smaller, (6.4 kJ/mol, but for the case (3) is smallest, about 4 kJ/mol). All configurations are the result of a weak H⋯F hydrogen bond with the interaction distance of the order of 2.7 – 3.5 Å. We observed a mesogen shift of one molecule by at least a third of the mesogen core length relative to the other interacting molecule, as well as a significant tilt of the molecules (Fig. 7).

For all systems the vibrational dipole derivatives have been calculated and compared with no interacting molecule. We noticed that the conformation of terphenyl did not significantly affect the cohesive energy of the H⋯F bond and we did not notice any substantial differences in the IR spectrum for the vibrations in range 800-1600 cm$^{-1}$. For this, the analysis of changes in the dipole transition moment is present on an example of the helical conformer.

We observed an increase of the perpendicular transition dipole, i.e. for bands at 905 cm$^{-1}$ and 1460 cm$^{-1}$ wavenumber by about 50%, and for parallel transition dipole (890 and 1485 cm$^{-1}$) a decrease of about 15% compared to a system without specific hydrogen bonds. This results well coincides with the experiment. Table 1 shows the dipole transition moment changes in the presence of hydrogen bonding and local orientational order for the case of arrangement (1) and (2). In the case (3) changes are not so significant and statistically scattered.

We have provided experimental and theoretical evidence of the stabilization of the $N_{TB}$ phase by arrays of multiple nonspecific short-range intermolecular interactions between the hydrogen atoms and the fluorine atoms substituted in the terphenyl ring, which as a high electron density area will act as a proton acceptor.

Table 1. Changes of the transition dipole moment: $\mu_{TB}^2/\mu_N^2$ in the presence of hydrogen bonding and local orientational order.

| DTC5Cn Dimer | $\mu_{TB}^2/\mu_N^2$ for 890 cm$^{-1}$ | | | $\mu_{TB}^2/\mu_N^2$ for 905 cm$^{-1}$ | | |
|---|---|---|---|---|---|---|
| | Experiment | DFT | | Experiment | DFT | |
| | | 1 pair H···F (1) | 2 pair H···F (2) | | 1 pair H···F | 2 pair H···F |
| DTC5C5 | 0.9$^a$ / 0.83$^b$ | 0.83 | 0.81 | 1.13$^a$ | 1.45 | 1.55 |
| DTC5C7 | 0.80 | | | 1.20 | | |
| DTC5C9 | 0.77 | | | 1.23 | | |
| DTC5C11 | 0.75 | | | 1.25 | | |
| | for 1485 cm$^{-1}$ | | | for 1460 cm$^{-1}$ | | |
| DTC5C5 | 0.88 | 0.84 | 0.82 | 1.13 | 1.23 | 1.33 |
| DTC5C7 | 0.81 | | | 1,19 | | |
| DTC5C9 | 0.78 | | | 1.20$_5$ | | |
| DTC5C11 | 0.77 | | | 1,23 | | |

$^a$ in $N_{TB}$ phase, $^b$ in SmA phase.

**Results summary**

All discussed above intermolecular interactions, must lead to the substantial bond order correlations in the $N_{TB}$ phase. The formation of the heliconical bond order locks the molecular twist-bend conformation, i.e., it makes the linkage effectively rigid. This suppresses the entropic penalty of packing the flexible linkage next to a rigid arm, therefore favoring positional disorder as was confirm by SAXS [5]. This means that the heliconical orientation is determined by the twist-bend molecular conformation, and does not require the positional correlations that would arise by "chaining" molecules to form a quasi-polymeric structure [33]. Therefore, the mechanism for the tight pitch of the twist-bend nematic liquid crystal phase is quite different from that of other systems such as proteins and DNA, which require both molecular chirality and more significant positional correlations that are generally absent in the $N_{TB}$ forming dimers.

An interesting result from simulation was obtaining a significant shift of interacting molecules for configurations representing the system of multiple hydrogen bonds (Fig.8). This can be achieved only in the arrangement of the cores significantly tilted with the respect of the helical axis. Notably, results of from SAXS, GISAX and resonant XRD indicate that a separation in local layers does not occur only between aromatic and aliphatic moieties, but also between the aliphatic spacers and aliphatic tails [2]. SAXS patterns of the N and $N_{TB}$ phases are very similar; this suggests the presence of local layering in the N phase as well, but there the conformations do not synchronize to form a helix.

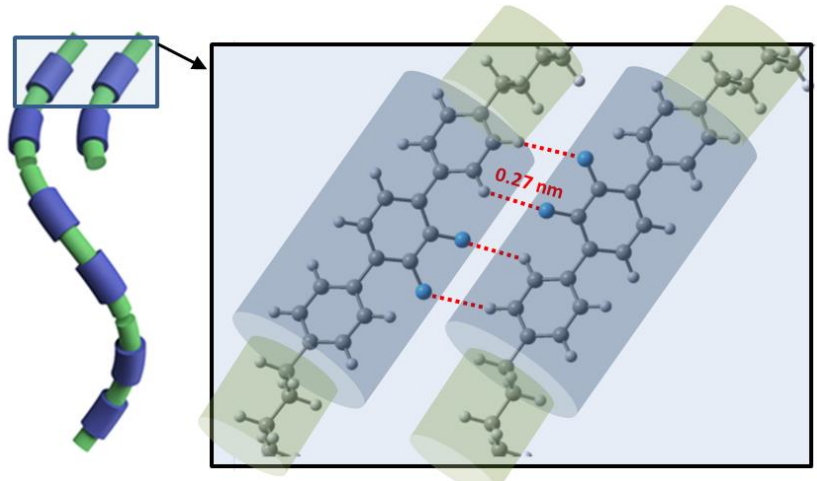

Fig. 8. Visualization of the $N_{TB}$ phase stabilization by arrays of multiple hydrogen bonds (XF····HX, X-benzene ring).

The proposed arrangement has also a strong influence on the dielectric properties in the $N_{TB}$ phase. As shown in Fig. 9 the measured data indicated that the electric dipoles of the difluorinated phenyl groups are arranged in an antiparallel manner. Thus, the total electric dipole of the assembly (molecular system) is considerably reduced in the $N_{TB}$ phase, Fig.9. Dielectric permittivity as an average, $(2\varepsilon_\perp+\varepsilon_{||}/3)$, refers to the square of the fixed dipole of the set per molecule. We can clearly see that this average shows a significant drop on entering the $N_{TB}$ phase. This is in accordance with results reported in the N. Sebastian et al [1].

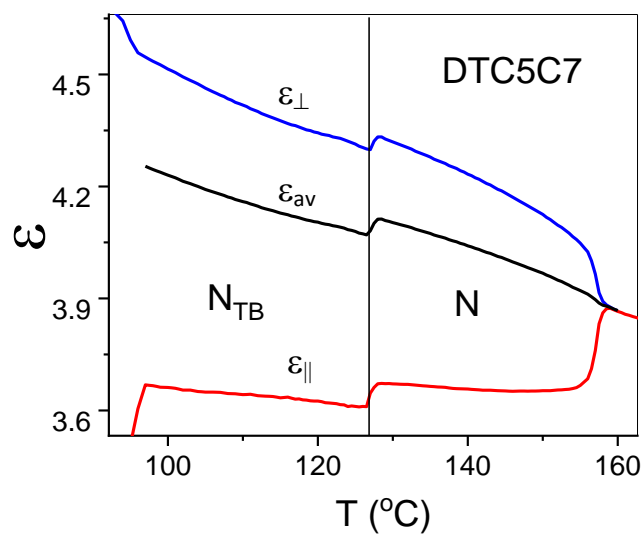

Fig.9. Temperature dependence of parallel and perpendicular components of the dielectric permittivity of the DTC5C7 dimer and its average dielectric permittivity.

Formation of the $N_{TB}$ phase, with short-range positional order and longer-range bond order, reflects a subtle interplay between the molecular bend and flexibility. The twist-bend nematic phase forms as the odd-membered linkage becomes sufficiently rigid along one axis to provide bend with angle $\beta$ between the rigid arms and a twist with angle $\alpha$ between the planes of the arms (propeller type structure)

**Conclusion**

The most important observation from spectroscopic and dielectric measurements is the sudden increase of the intermolecular interactions as the temperature decrease from the nematic to the twist-bend phase. This is demonstrated by significant increase of correlations of the induced dipoles in the dimer cores. The longitudinal induced dipoles of the core show negative correlations (antiparallel order) while lateral ones are positively correlated. Dielectric data also confirm correlations of the permanent dipoles. These results well corresponds with a significant grow of the molecular biaxiality in the $N_{TB}$ phase transition in contrast the case of the *N* phase, where such a biaxiality is absent. Several self-assembly of dimers pairs were analyzed using DFT calculation by optimizing their energy. A number of dimer arrangement showing hydrogen H···F bonds were identified that leading to significant bond order.

**References:**

[1] N. Sebastián, M.G. Tamba, R. Stannarius, M. R. de la Fuente, M. Salamonczyk, G. Cukrov, J. Gleeson, S. Sprunt, A. Jákli, C. Welch, Z. Ahmed, G.H. Mehl and A. Eremin, *Phys. Chem. Chem. Phys.*, 2016, **18**, 19299.

[2] W.D. Stevenson, Z. Ahmed, X.B. Zeng, C. Welch, G. Ungar and G.H. Mehl, *Phys. Chem. Chem. Phys.,* 2017, **19**, 13449.

[3] W.D. Stevenson, H-X Zou, X-B Zeng, Ch. Welch, G. Ungar and G.H. Mehl, *Phys. Chem. Chem. Phys*., 2018, **20**, 25268.

[4] R. Saha, G. Babakhanova, Z. Parsouzi, M. Rajabi, P. Gyawali, Ch. Welch, G.H. Mehl, J. Gleeson, O.D. Lavrentovich, S. Sprunt and A. Jákli, *Mater. Horiz*., 2019, **6**, 1905.

[5] R. Saha, C. Feng, C. Welch, G.H. Mehl, J. Feng, C. Zhu, J. Gleeson, S. Sprunt and A. Jákli, 2020, arXiv:2011.09535v1 [cond-mat.soft].

[6] S.M. Jansze, A. Martinez-Felipe, J.M.D. Storey, A.T.M. Marcelis, and C.T. Imrie, *Angew. Chem. Int. Ed*. 2015, **54**, 643.

[7] R. Walker, D. Pociecha, J.P. Abberley, A. Martinez-Felipec, D.A. Paterson, E. Forsyth, G.B. Lawrence, P.A. Henderson, J.M.D. Storey, E. Gorecka and C.T. Imrie, *Chem. Comm*. 2018, **54**, 3383.

[8] M. Alaasar & C. Tschierske, *Liq. Cryst*. 2019, **49**, 124

[9] A. Martinez-Felipe, A.G. Cook, J.P. Abberley, R. Walker, J.M. D. Storey and C.T. Imrie, *RSC Adv.*, 2016, **6**, 108164.

[10] A. Martinez-Felipe, F. Brebner, D. Zaton, A. Concellon, S. Ahmadi, M. Piñol, L. Oriol, *Molecules,* 2018, **23**(9), 2278.

[11] D.A. Paterson, A. Martínez-Felipe, S.M. Jansze, A.T.M. Marcelis, J.M.D. Storey & C.T. Imrie, *Liq. Cryst*. 2015, **42** (5-6), 928.

[12] R. Paulini, K. Müller, F. Diederich, *Angew. Chem. Int. Ed.,* 2005, **44**, 1788.

[13] G.T. Giuffredi, V. Gouverneur, B. Bernet, *Angew. Chem. Int. Ed.*, 2013, **52**, 10524.

[14] V.K. Batra, L.C. Pedersen, W.A. Beard, S.H. Wilson, B.A. Kashemirov, T.G. Upton, M.F. Goodman, C.E. McKenna, *J. Am. Chem. Soc*., 2010, **132** (22), 7617.

[15] C. Dalvit, C. Invernizzi, and A. Vulpetti, *Chem. Eur. J.,* 2014, **20**, 11058.

[16] C. Ouvrard, M. Berthelot, C. Laurence, *J. Phys. Org. Chem*., 2001, **14**, 804.

[17] J.A.K. Howard, V.J. Hoy, D. O'Hagan, G.T. Smith, *Tetrahedron*, 1996, **52**, 12613.

[18] R.E. Rosenberg, *J. Phys. Chem. A*, 2012, **116** (40), 10842.

[19] K. Müller, C. Faeh, F. Diederich, *Science*, 2007, **317** (5846), 1881.

[20] A. Vulpetti, C. Dalvit, *Drug Discovery Today*, 2012, **17**, 890.

[21] E.B. Wilson, J.C. Decius, and P.C. Cross, *Molecular Vibrations*, McGraw-Hill, New York, 1955.

[22] P.G. deGennes, J. Prost, The Physics of Liquid Crystals, Oxford Science Publications, second edition, 1993.

[23] D. Dunmur, K. Toriyama, in Handbook of Liquid Crystals edited by D. Demus et al, Chapter VII, Vol. 1A, 189, 2001.

[24] Gaussian 09, Revision E.01, Frisch, M.J., Trucks, G.W., Schlegel, H.B., et al.

[25] a) A.D. Becke, " Density-functional exchange-energy approximation with correct asymptotic behaviour" Phys. Rev. A, 38, 3098 (1988), doi:10.1103/PhysRevA.38.3098.

b) R.H. Hertwig, W. Koch " On the parameterization of the local correlation functional. What is Becke-3-LYP?" Chem. Phys. Lett., 268 (5–6), 345-351 (1997). Doi:10.1016/S0009-2614(97)00207-8.

[26] J.B. Foresman and A. E. Frisch, Exploring Chemistry with Electronic Structure Methods Gaussian, Pittsburgh, 1998.

[27] J.W. Emsley, G. De Luca, A. Lesage, D. Merlet & G. Pileio, *Liq Cryst.* 2007, **34**(9), 1071.

[28] K. Merkel, B. Loska, Ch. Welch, G.H. Mehl, A. Kocot, 2020, arXiv:2010.14896 [cond-mat.soft]

[29] R.J. Mandle, E.J. Davis, C.T. Archbold, C.C.A. Voll, J.L. Andrews, S.J. Cowling, J.W. Goodby, *Chem. Eur. J.,* 2015,**21**, 8158.

[30] H.B. Jansen and P. Ross, *Chem. Phys. Lett*. 1969, **3**, 140.

[31] S.B. Boys and F. Bernardi, *Mol. Phys*. 1970, **19**, 553.

[32] K. Merkel, A. Kocot, R. Wrzalik, J. Ziolo, *J. Chem. Phys*., 2008, **129**, 074503.

[33] F.F.P Simpson, R.J. Mandle, J.N. Moore, J.W. Goodby, J. Mater. Chem. C, 2017, **5** (21), 5102.

[34] A.G Vanakaras, D.J. Photinos, Soft Matter 2016, **12** (12), 2208.

## Conflicts of interest

There are no conflicts to declare.

## Acknowledgements

Authors (KM & AK) thank through the National Science Centre, Poland for Grant No. 2018/31/B/ST3/03609. All DFT calculations were carried out with the Gaussian09 program using the PL-Grid Infrastructure on the ZEUS and Prometheus cluster.